\documentclass[sn-mathphys,Numbered]{sn-jnl}
\usepackage{graphicx}%
\usepackage{multirow}%
\usepackage{amsmath,amssymb,amsfonts}%
\usepackage{amsthm}%
\usepackage{mathrsfs}%
\usepackage[title]{appendix}%
\usepackage{xcolor}%
\usepackage{textcomp}%
\usepackage{manyfoot}%
\usepackage{booktabs}%
\usepackage[ruled,vlined]{algorithm2e}
\usepackage{listings}%

\usepackage{braket}
\usepackage{url}
\usepackage{color,soul}
\definecolor{darkBlue}{RGB}{0,0,130}
\definecolor{darkGreen}{RGB}{13, 59, 2}

\begin{document}

\title[Article Title]{Machine Learning Catalysis of quantum tunneling 
}

\author[1]{\fnm{Renzo} \sur{Testa}}
\email{renzo.testa@yahoo.com}

\author[1,4]{\fnm{Alex} \sur{Rodriguez}}
\email{alejandro.rodriguezgarcia@units.it}

\author[1]{\fnm{Alberto} \sur{d{'}Onofrio}}
\email{alberto.donofrio@units.it}

\author[2,3]{\fnm{Andrea} \sur{Trombettoni}}
\email{atrombettoni@units.it}

\author[2]{\fnm{Fabio} \sur{Benatti}}
\email{fabio.benatti@units.it}

\author[1,5]{\fnm{Fabio} \sur{Anselmi}}
\email{fabio.anselmi@units.it}

\affil[1]{\orgdiv{Department of Mathematics Informatics and Geoscience}, \orgname{University of Trieste}, \orgaddress{\street{Via Alfonso Valerio 2}, \city{Trieste}, \postcode{34127}, \country{Italy}}}

\affil[2]{\orgdiv{Department of Physics}, \orgname{University of Trieste}, \orgaddress{\street{Strada Costiera 11}, \city{Trieste}, \postcode{I-34151}, \country{Italy}}}

\affil[3]{ \orgname{SISSA and INFN}, \orgaddress{\street{Via Bonomea 265}, \city{Trieste}, \postcode{I-34136}, \country{Italy}}}

\affil[4]{ \orgname{ICTP}, \orgaddress{\street{Strada Costiera 11}, \city{Trieste}, \postcode{I-34151}, \country{Italy}}}

\affil[5]{\orgname{MIT}, \orgaddress{\street{77 Massachusetts Ave}, \city{Cambridge}, \postcode{02139}, \country{MA,USA}}}

\abstract{
Optimizing the probability of quantum tunneling between two states, while keeping the resources of the underlying physical system constant, is a task of key importance due to its critical role in various applications.\\
We show that, by applying Machine Learning techniques when the system is coupled to an ancilla, one optimizes the parameters of both the ancillary component and the coupling, ultimately resulting in the maximization of the tunneling probability.\\
We provide illustrative examples for the paradigmatic scenario involving a two-mode system and a two-mode ancilla in the presence of several interacting particles. 
Physically, the increase of the tunneling probability is rooted in the decrease of the two-well asymmetry due to the coherent oscillations induced by the coupling to the ancilla.\\
We also argue that the enhancement of the tunneling probability is not hampered by weak coupling to noisy 
environments.}
\maketitle


\vskip 1cm

Quantum tunneling is a distinctive property of quantum mechanics \cite{Roy86} that plays a crucial role in many physical processes,  
from chemical reactivity \cite{chemical_tunneling} and nuclear fusion \cite{nuclear_fusion} to the alpha-radioactive decay of atomic nuclei \cite{Tunneling_radioactive}. It also has a myriad of diverse technological applications, such as in tunnel diodes \cite{tunnel_diodes} 
in scanning tunneling microscopes \cite{tunneling_microscope} and 
in programming the floating gates of flash memories \cite{flash_memories}.
A paradigmatic application of quantum tunneling is the Josephson effect~\cite{Barone82} which finds several practical applications in high-precision measurements of voltage and magnetic fields 
\cite{Tinkham96} as well as in superconducting qubits for quantum computing \cite{Makhlin01,Google2017}. 
A general challenge in this context is therefore to be able to control quantum tunneling in a tunable way.
We start considering a single particle within a double-well potential. This model has not only an academic meaning, as it also exhibits the main features of more complex macroscopic tunneling phenomena which include Josephson tunneling between two superconductors \cite{Barone82} and tunneling between two Bose-Einstein condensates \cite{Javanainen86,Smerzi97}. 
Specifically, the two-mode model involves systems with  ({\it a}) well-separated energy levels and ({\it b}) coupling between these levels determining the two states between which tunneling occurs.
Well-known concrete realizations are given by the transitions left-right in a spatial double-well trap, between up and down spin states, or between $\ket{0} $ and $\ket{1}$ states in quantum bit scenarios \cite{Nielsen10}.
Furthermore, the intricate influence of the interactions between the target two-level systems and external ones must be taken into account \cite{Kagan92,Zwerger87}.

In this context, a natural question arises: can we inverse-design the coupled ancillary systems and their interactions with the target system to amplify the quantum tunneling probability of the latter?

Different methods to enhance the probability have been studied such as chaos-assisted tunneling \cite{Chaos_Assisted} and resonance tunneling \cite{AssRes} or tailoring the energy barrier between the two states, \cite{Kagan91},\cite{Kagan92}. 
Here, we explore a different approach by studying the interaction of a quantum system with a similar-type ancilla system with a generic coupling.  
Ancillary-assisted protocols have been employed in a variety of tasks such as quantum tomography \cite{assisted_tomography}, copying of quantum states \cite{local_copying} and metrology \cite{Metrology}.
In the case of quantum tunneling, to the authors' knowledge, the optimization by automatic differentiation of such protocols in the presence of several particles interacting among themselves and with their environment have not yet been considered. The novelty and purpose of our approach consist in the ability to optimize by machine learning techniques: {\it 1)} the parameters of the ancilla Hamiltonian and its initial state; {\it 2)} the form of the ancilla-system coupling.

We start by analyzing the simpler example of a single boson interacting with $N_A$ ancillary trapped bosons described by a particular two-site Bose-Hubbard Hamiltonian, a widely used model in the study of interacting particles~\cite{Smerzi97,Jaksch97,BE_condensation1}. 
Specifically, we show how to enhance the tunneling probability from left to right of a system composed of $N_S$ trapped bosons by coupling them with a learned ancillary double-well system. In the context of quantum computation, this example can be viewed as a transition from state $0$ to state $1$, or equivalently, as a NOT-gate operation \cite{Nielsen10}.
Thus, enhancing left-to-right tunneling probability becomes crucial, especially when facing potential noise that might diminish the required quantum coherence for tunneling.

In the following, starting from the the analytically solvable scenarios, we conduct comprehensive simulations including systems and ancillas with many particles, the effects of interactions and of the presence of a decohering noisy environment~\cite{BRE02} affecting both the system and the coupled ancilla.
The simulation outcomes demonstrate the efficacy of coupling a tunneling quantum system with an inverse-designed ancilla to enhance tunneling probability. 

\section*{Trapped two-mode bosonic systems}

In the following we focus on a quantum systems $S$ consisting of $N_S$ interacting bosons trapped in a double-well potential~\cite{Smerzi97,BE_condensation1}.

In the absence of couplings to external systems or to an environment, the trapped bosons evolve in time according to a two-mode Bose-Hubbard Hamiltonian, which, using the Jordan-Schwinger representation, reads:
\begin{equation}
\label{HamS}
H_S=\eta_S\, J^{2}_{z}\, -\, \gamma_S\,J_{x} - \Delta_S\, J_{z}\   
\end{equation}
(see Methods for the definition of the $J_{x,y,z}$ matrices). 
The real coefficient $\gamma_S$ governs the energy barrier's height, driving tunneling between the left and right wells, $\Delta_S$ introduces energy asymmetry, while $\eta_S$ quantifies repulsion ($\eta_S>0$) or attraction ($\eta_S<0$) among bosons and  is proportional to the $s$-wave scattering length \cite{BE_condensation2}. An estimate of the parameters $\gamma_S, \Delta_S$ 
gives $\gamma_S /k_B\sim 0.5 nK$ and $\Delta_S / k_B \sim 0.2 nK$, while $\eta_S$ can be varied in the interval $\sim k_B \cdot [0.1-10 nK]$ (see the Supplementary Material).
The number-preserving, reversible dynamics for the states (density matrices) $\rho^{(S)}$ of the $N_S$ trapped bosons is thus generated by the master equation  $\partial_t\rho^{(S)}(t)=-i{[H_S\,,\,\rho^{(S)}(t)]\,}$.

When all bosons are initially confined in the left well at time $t=0$, their state is $\rho^{(S)}_L=\ket{N_S}\bra{N_S}$, where $\ket{k}$ describes the state with $k$ boson in the left well and $N_S-k$ in the right one and satisfies $J_z\ket{k}=(N_S-2k)\ket{k}$.
The states $\rho^{(S)}_L$ might be used to encode the logical qubit state $\ket{0}$ which then evolve into the logical qubit state $\ket{1}$ when all $N_S$ bosons have tunneled to the right well turning the initial boson state into $\rho^{(S)}_R=\ket{0}\bra{0}$. Within this setting, one is interested in maximizing the tunneling probability $P_{L\to R}(t_*)$ of the $N_{S}$ bosons from the left to the right well at a certain time $t^{*}$. 
Such a probability is given by
\begin{equation}
\label{eq:probability}
P_{L\to R}(t) = {\rm Tr}(\rho^{(S)}_{L}(t)\rho^{(S)}_{R}) =\langle 0\vert  \rho^{(S)}_{L}(t)\vert 0\rangle\ ,
\end{equation}
where 
\begin{equation}
\label{eq:sol}
\rho^{(S)}_L(t) = \exp\left(-\frac{i}{\hbar}\,H_S\,t\right)\,\rho^{(S)}_L\, \exp\left(+\frac{i}{\hbar}\,H_S\,t\right)\ .
\end{equation}

\subsection*{Two-level systems}

For $N_S=1$, a two-mode $N_S$-body system trapped by a double-well potential reduces to a two-level system described by the Hamiltonian 
$H_S=-\Delta\,\sigma_{z}-\gamma\,\sigma_{x}$, where $\sigma_{x,z}$ are the Pauli matrices, while $J_z^2=\sigma_z^2/4=I/4$ in~\eqref{HamS} can be neglected.
The states with one particle in the left and right wells are the eigenstates $\ket{L}$ and $\ket{R}$ of $\sigma_z$, corresponding to eigenvalues $-1$ and $+1$.
Then, (see Methods) the left-to-right tunneling probability at time $t$ is
\begin{equation}
    \label{1partunn}
P_{L\to R}(t)=\left\vert\langle R\vert L\rangle_t\right\vert^2=\frac{\gamma^2}{\hbar^2\omega^2}\,\sin^2(\omega t)
=\frac{\sin^2(\omega t)}{1+\frac{\Delta^2}{\gamma^2}}\ ,\qquad \hbar\,\omega=\sqrt{\Delta^2+\gamma^2}\ .
\end{equation}
Thus, the tunneling probability can never reach its maximum $P_{L\to R}(t)=1$ for asymmetric traps with $\Delta\neq 0$.

\section*{Noiseless coupling to ancilla systems}

Here we consider two double-well potentials and a density-density interaction among the bosons trapped by them. 

The first double-well potential, containing $N_S$ bosons and characterized by a fixed Bose-Hubbard Hamiltonian, functions as the target system denoted as "S." The second double-well potential serves as a trainable ancillary system denoted as "A," and it accommodates $N_A$ trapped bosons. The ancillary system is governed by the Hamiltonian  in~\eqref{HamS}, with adjustable energy parameters $\eta_{A},\gamma_{A},\Delta_{A}$.
We consider a density-density interaction $H_{int}=\alpha\, J_{z}\otimes J_{z}$ between the bosons of the two double-wells, with interaction strength  $\alpha$, also learnable. This interaction could be experimentally implemented using e.g. dipolar or Rydberg atoms \cite{Defenu21}.

In the following, we distinguish two different physical contexts: the first one considers asymmetric traps, for which reaching certainty of left-to-right tunneling is impossible, and see whether probability $1$ can instead be reached by coupling to ancillas. In this case, we measure energies in units of the system energy asymmetry $\Delta_S$ and time in units of $\hbar/\Delta_S$, equivalently setting $\hbar=\Delta_S=1$.

The second one considers traps with vanishing energy asymmetry and see whether,
through a suitable coupling to ancillas, one can reach perfect tunnelling of all system bosons from left to right in a shorter time with respect to the case without ancillas. In such a case, $\Delta_S=0$ and we measure energies in unit of $\gamma_S$ and time in units of $\hbar/\gamma_S$, equivalently setting $\hbar=\gamma_S=1$. In both cases, the time unit for a system of coupled ultracold dipolar gases (see Supplementary Material) is of the order of $10 ms$.

Starting from an initial state of the tensorized form $\rho^{(S)}_L\otimes\rho^{(A)}$
where $\rho^{(S)}_L=\ket{N_S}\bra{N_S}$ denotes the initial state with all $N_S$ bosons of the system of interest trapped in the left well, while $\rho^{(A)}$ denotes any initial state of the ancillary double-well potential, the system state at time $t$ is then given by
\begin{equation}
\label{red-dyn}
\rho^{(S)}_{L}(t) = {\rm Tr}_{A}\Big(e^{-i\,H_{SA}\,t}\,\rho^{(S)}_{L}\otimes \rho^{(A)} e^{i\,H_{SA}\,t}\Big)\ ,  \end{equation}
where $H_{SA}=H_S+H_A+H_{int}$ is the total Hamiltonian of the compound system $S+A$ and ${\rm Tr}_A$ denotes the partial trace with respect to the ancillary system.

We apply machine learning techniques to optimize the coupling strength $\alpha$, the energy parameters $\eta_A,\Delta_A,\gamma_A$ together with the initial state $\rho^{(A)}$ and the time $t$.
All these parameters will be learned in such a way to maximal transfer probability $P_{L\to R}(t) = \langle 0\vert\rho^{(S)}_{L}(t)\vert 0\rangle$ of the $N_S$ bosons of the target system $S$ from the left to the right well.

\section*{Two simple limit cases}

As first benchmark example, consider a one-boson target system, $N_S=1$, in an asymmetric couple to $N_A \equiv N$ non-interacting and non-tunneling trapped bosons: 
\begin{equation}\label{2qubits}
H_{S} =-\sigma_z-\gamma\,\sigma_x\,\quad H_A=-\,J_z\,\quad H_{int}=\alpha\,\sigma_z\otimes J_z\
\end{equation}
where, according to the assumed convention of measuring energies in units of $\Delta_S$, we set $\Delta_S=\Delta_A=1$.
By proceeding as in the Supplementary Material we consider the joint initial state of system and ancilla to be a tensor product state $\rho^{(S)}_L\otimes\rho^{(A)}$ corresponding to the system with its single boson localized in the left well, $\rho^{(S)}_L=\ket{N} \bra{N}$ and to the ancilla being in a generic pure state $\rho^{(A)}=\ket{\psi_A}\bra{\psi_A}$ of its $N$ bosons. 
Then, the left-to-right transition probability at $t\geq 0$, 
\begin{equation}
    \label{2qubitprob2}
P_{L\to R}(t) = \sum_{k=0}^N\Big|\langle k\vert\psi_A\rangle\Big|^2\,\frac{\gamma^2}{\omega_k^2}\,\sin^2(\omega_k t)\ ,\quad\omega_k:=\sqrt{\left(1-\alpha\frac{N-2k}{2}\right)^2\,+\,\gamma^2}\ ,
\end{equation}
can beat the upper bound holding for $N_S=1$. Notice that unit probability can be reached at time $t^{*}=(\pi/2+n\pi)/\gamma$ only if $\ket{\psi_A}$ is an eigenstate of $J_z$ and the coupling $\alpha$ is such that the corresponding energy asymmetry $1-\alpha\frac{N-2k}{2}$ vanishes.

The physical mechanism behind the increase of the tunneling probability can be read from~\eqref{2qubitprob2}.
Indeed, setting $k=N$ and comparing it with~\eqref{1partunn} where $\Delta=1$, one sees that the coherent oscillations induced by the coupling to the ancilla redefine an effective two-well asymmetry which vanishes  for a suitable interaction strength $\alpha$.

As a second benchmark example, consider one boson trapped in a symmetric trap, coupled to an asymmetric and non-tunneling ancillas via the following Hamiltonian:
\begin{equation}
    \label{2qubitssymm}
    H_S=-\sigma_x\ ,\quad H_A=-\,J_z\ ,\quad H_{int}=\alpha\,\sigma_x\otimes J_z\ ,
\end{equation}
where we measure energies in units of $\gamma_S$ and set $\gamma_S=1$, $\gamma_A=0$ and $\Delta_A=0$.

As shown in the Supplementary Material, the left-to-right tunneling probability of the single trapped boson is given by
\begin{equation}
    \label{2qubitprob2sym}
P_{L\to R}(t) = \sum_{k=0}^N\Big|\langle k\vert\psi_A\rangle\Big|^2\,\sin^2(\omega_k t)\ ,\quad\omega_k:=\left\vert 1-\alpha\frac{N-2k}{2}\right\vert\ .
\end{equation}
Unit tunneling probability now occurs, for $\ket{\psi_A}=\ket{k}$, at  $t_k=\pi/(2\omega_k)$, in general faster than without coupling to ancillas ($\alpha=0$).

\section*{Decoherence}

In the presence of many particles, noise and dissipation, hence decoherence, cannot be easily neglected. 
We thus now consider the optimization of tunneling when both system $S$ and ancilla $A$ cannot be considered isolated from their environment even if they interact very weakly with it. In such a case, one expects decoherence to suppress tunneling. This effect is best described by considering both system $S$ and ancilla $A$ independently undergoing a reduced irreversible dynamics 
generated by a master equation of the so-called Gorini-Kossakowski-Sudarshan-Lindblad (GKSL) form~\cite{BRE02}:
\begin{equation}
\label{GKSL}
\partial_t\rho(t)=-i[H\,,\,\rho(t)]\,+\,\lambda\,\left(J_z\,\rho(t)\,J_z-\frac{1}{2}\left\{J_z^2\,,\,\rho(t)\right\}\right)\
\end{equation}
where $\{X\,,\,Y\}$ denotes the anti-commutator and $\lambda$ is a small, dimensionless constant that measures the strength of the interaction between the system and the environment.

\section*{Numerics for a 1-boson system}

In order to test the learning procedure, we use the simplest analytically solvable case with $N_S=N_A=1$; namely, the case of one non self-interacting ($\eta_S=0$), tunneling ($\gamma_S\neq 0$) boson with energy asymmetry $\Delta_S=1$, coupled to an ancillary
double-well potential also trapping a single non-self-interacting and non-tunneling boson ($\eta_A=\gamma_A=0$) with same fixed energy asymmetry, $\Delta_A=\Delta_S=1$. 

The plots shown in Figure~\ref{fig:test1} (and Figure~\ref{fig:test2} in the Supplementary Material) provide clear evidence that the learning method rapidly converges to the optimal parameter values that maximize the tunneling probability
while the system without the ancilla reaches a maximum probability of $0.2$.
In the simple case considered Figure~\ref{fig:test1} (Top Left) the learning algorithm optimizes the coupling $\alpha$ between the two double-well traps which, in its turn, determines the tunneling probability through~\eqref{2qubitprob2} with $N=1$ (\textit{Bottom}). 
Specifically, the model consistently learns the optimal value $\alpha=-2$, as expected for maximizing~\eqref{2qubitprob2} when $N=1$, and the ancilla is initialized to $| \Psi_A \rangle = |L \rangle$. Similarly a value of $\alpha=+2$ is learned when the ancilla is initialized to $| \Psi_A \rangle = |R \rangle$ (not shown).
We also report the time $t$ needed to reach the maximal probability: as expected from the theory $t^{*}\approx \pi$. Interestingly, in the case when the max probability reached by the system is already one, adding a time constraint in the optimization (see Methods) decrease the time to reach the maximal probability (Top Right) in agreement with eq. \eqref{2qubitprob2sym}.
In the next section, where a more general (not analytically solvable) setting will be considered,  the state of the ancilla will not be fixed; rather, it will be part of the optimization process.

\begin{figure}[t]
\centering
\includegraphics[width=1.0\linewidth]{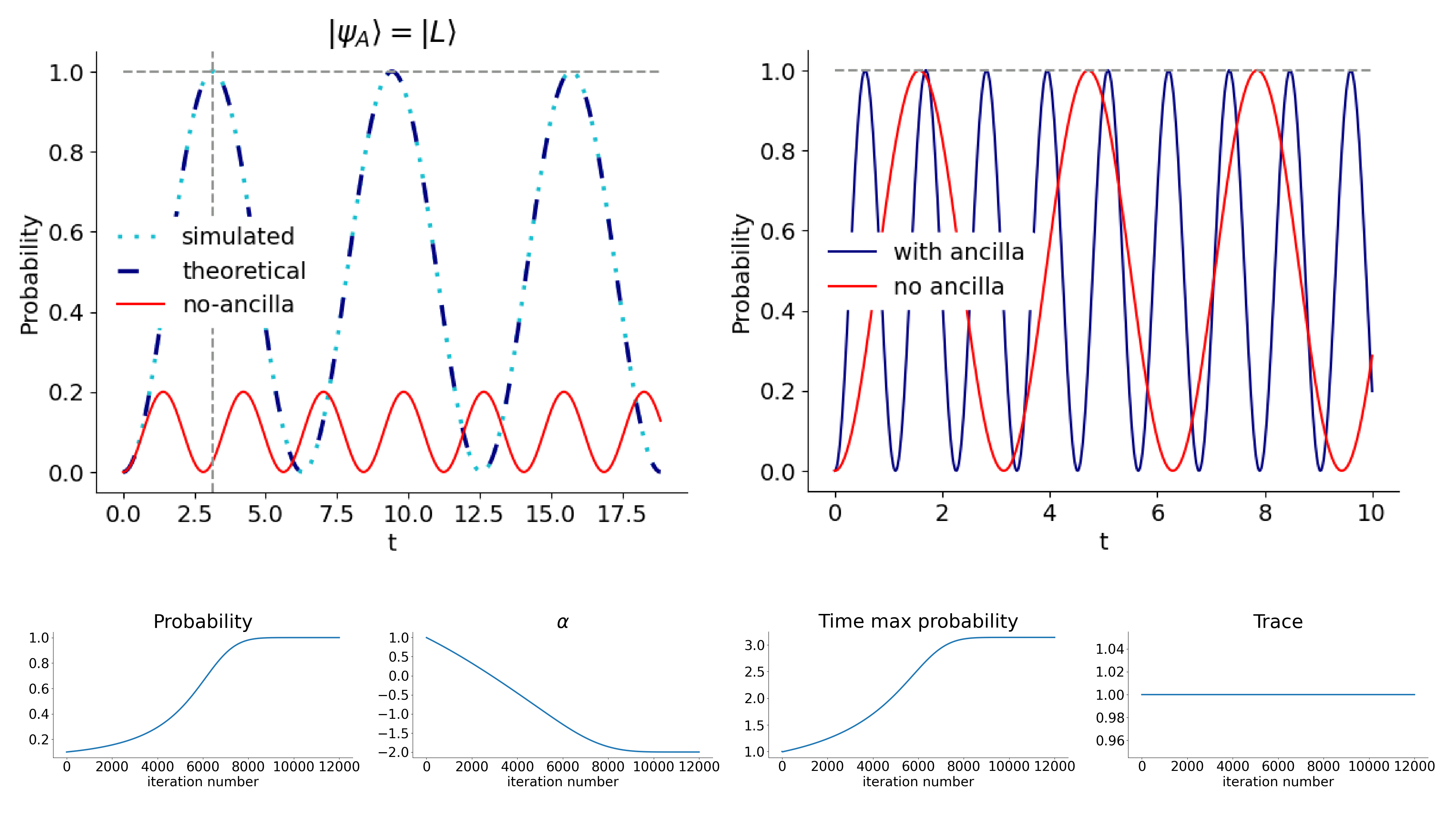}
\caption{ \textit{Top Left}: After learning. Time-evolution of the tunneling probability for the system coupled with the ancilla compared to the system without the ancilla, reporting both the theoretical predictions and the 
optimized results. \textit{Top Right}: After learning in the case when the system can reach a unit probability and we minimize the time:the maximal tunneling probability is reached earlier when the ancilla is coupled to the system. 
In both cases time is reported in $\hbar/\Delta_{S}$ units, each corresponding to $\sim 10 ms$.
\textit{Bottom}: Evolution, during optimization, of the tunneling probability  and the learnable parameters $\alpha$ (interaction strength) and of the time to reach the maximum probability (for the figure Top Left). Also, we report the trace of the learned ancilla as a sanity check. 
} 
\label{fig:test1}
\end{figure}

\section*{Multi-particle noiseless tunneling and physical resources }

The noiseless case, where the number of bosons grows in both the double-well system and ancilla, represents the next natural step in  increasing model complexity. 
It also functions as a reference to explore the relationships among the learnable parameters governing the phenomenon. 

\begin{figure}[t]
\centering
\begin{minipage}{.45\textwidth}
  \centering
  \includegraphics[width=1.0\linewidth]{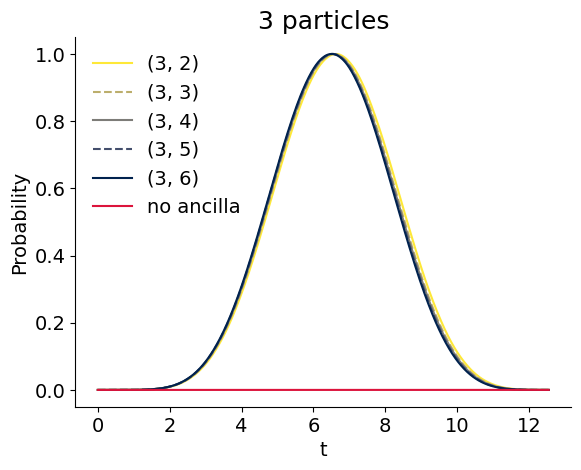}
\end{minipage}%
\begin{minipage}{.45\textwidth}
  \centering
  \includegraphics[width=1.0\linewidth]{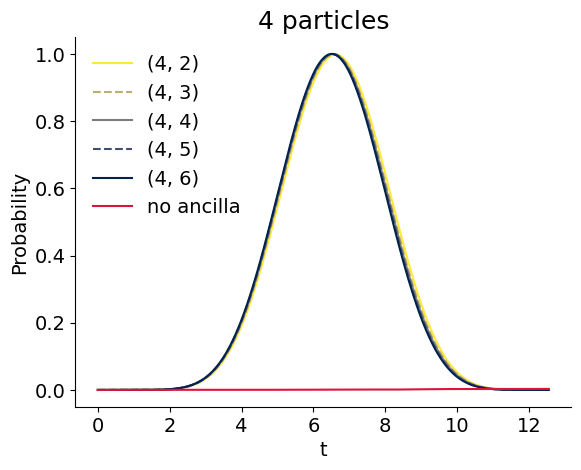}
\end{minipage}
\caption{Time evolution of tunneling probability for $(N_S,N_A)$ particles in the system and ancilla with $N_S=3,4$ and $N_A=2,\cdots,6$ compared with that of the system with no coupling (red line).}
\label{fig:test34}
\end{figure}

Figure \ref{fig:test34} confirms that, despite the tunneling probability of $N_S$ bosons from the left to the right well being vanishingly small in the absence of coupling to an ancilla, it is nevertheless possible to achieve $P_{L \rightarrow R}^{max} = 1$ for various configurations involving different sizes of the system and ancilla, by learning the interaction  between them and the initial state of the ancilla.  

Interpreting the learnable parameters $\eta_{A},\gamma_{A},\Delta_{A},\alpha$ as learning resources 
a natural question is how they behave varying the number of bosons in the system and ancilla. We refer to the Methods and the Supplementary Material for a discussion of a realistic set-up in which such parameters can be tuned.

\begin{figure}[t]
\centering
\includegraphics[width=0.6\linewidth]{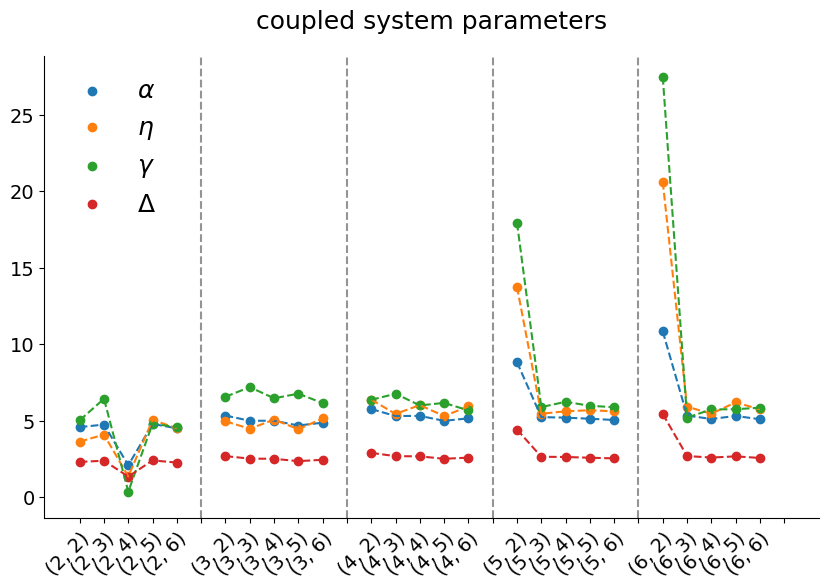}
\caption{ Variability of the learned parameters in function of the couples $(N_S,N_A)$ reporting system and ancilla particle numbers.}
\label{fig:test3}
\end{figure}

The plots of Figure \ref{fig:test3} indicate a decrease of $\eta_{A},\gamma_{A},\Delta_{A},\alpha$ with increasing ancilla size. 
Specifically, it appears that, for a given system, an ancilla with larger numbers of bosons 
requires weaker coupling to achieve the maximum tunneling probability. The exact or estimated relation between the maximum reachable tunneling probability and the sizes of the system and ancilla
will be a matter of future research.

\section*{Decoherence: a proof of concept}

We now show that the method developed in the previous sections is, as a matter of principle, applicable even in the presence of decoherence without severely affecting the ancilla parameters and initial state so to make them physically unattainable.

Indeed, let us consider the dissipative open dynamics of eq. \eqref{GKSL} which asymptotically lead to equally distributed mixtures of $0\leq k\leq N$ bosons localized in the left well and $N-k$ in the right one (see Methods). 
As a consequence of the coupling to an ancilla independently affected by decoherence, the time-evolution of the joint initial state of the system and ancilla is given by the following master equation: 
\begin{eqnarray}
\nonumber
    \partial_{t}\rho^{(SA)}(t) = &-&i[H_{SA},\rho^{(SA)}(t)]\\
    \nonumber
    &+&\lambda_A\Big((I\otimes J_z)\,\rho^{(SA)}(t)\,(I\otimes J_z)-\frac{1}{2}\{I\otimes J_z^2\,,\,\rho^{(SA)}(t)\}\Big)\\
    \label{eq:Noisy}
    &+& \lambda_S\Big((J_z\otimes I)\, \rho^{(SA)}(t)\, (J_z\otimes I)-\frac{1}{2}\{J_z^2\otimes I\,,\,\rho^{(SA)}(t)\}\Big)\ ,
\end{eqnarray}
where $\lambda_S,\lambda_A$ positive constants determine the noise strength which, for our simulations, have been fixed,in the spirit of the so-called weak-coupling limit,  to a value of $0.01$.

\begin{figure}[t]
\centering
\begin{minipage}{.5\textwidth}
  \centering
\includegraphics[width=1.0\linewidth]{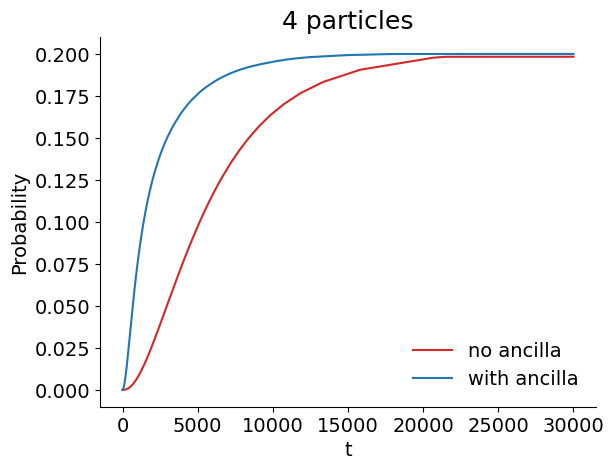}
\end{minipage}%
\begin{minipage}{.5\textwidth}
  \centering
  \includegraphics[width=1.0\linewidth]{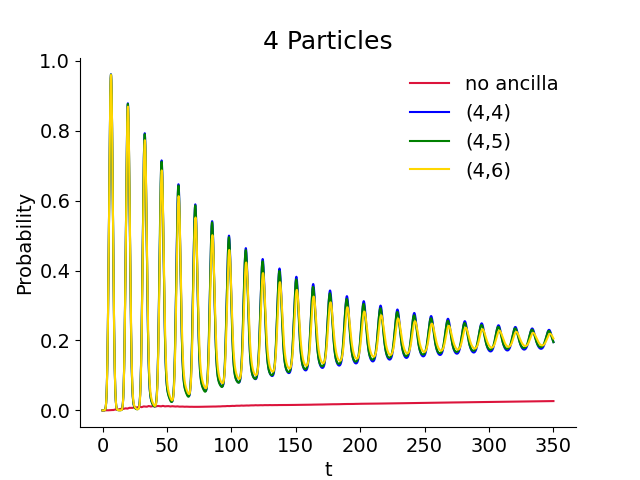}
\end{minipage}
\caption{\textit{Left:} The asymptotic behavior of the tunneling probability for a 4 particles system (red) and system coupled with ancilla (blue) with the same number of particles (random fixed parameters, no learning). \textit{Right:} Tunneling probability evolution after optimization for a system of $4$ particles coupled with an ancilla of $4,5,6$ particles. Red line: system with no coupling. Note the difference in the time scales between the right and left plots. Both system and ancilla are initialized, before learning, to the state where all particles are in the left well.}
\label{fig:B34}
\end{figure}

Figure \ref{fig:B34} (Left) reports the evolution of the tunneling probability where the learnable parameters are in this case randomly chosen. We note that, in agreement with eq.~\eqref{limitst}, the asymptotic state reaches a maximal tunneling probability of $0.2$. This is the case both for a system of $4$ particles without ancilla and a system plus ancilla both with $4$ particles.   
Figure \ref{fig:B34} (Right) reports
the evolution of the tunneling probability in time for $N_S=4$  with $N_A=4,5,6$ particles.
Interestingly, we note that, although the same asymptotic probability is attained, when we optimize the parameters: (a) the asymptotic state probability is reached within orders of magnitude faster and (b) for a brief period, a significantly higher probability can be achieved compared to the non-optimized scenarios.
To increase the experimental implementability of the learned ancilla states we also imposed them to be diagonal in the well-occupation number states.
Results are qualitatively similar to those reported in Figure \ref{fig:B34} (see Supplementary Material)

\section*{Outlook}

The idea of engineering an ancillary system and its coupling to a specific quantum system to enhance particular properties extends beyond merely improving tunneling probability. It introduces a comprehensive framework with diverse applications, in the same spirit of \cite{Hlearning,OpenSML}. 
In this work we addressed and solved the issue whether it is possible to learn an ancillary system and its interaction with the target system in such a way as to 
optimize the system tunneling properties.

Another significant area of research involves enhancing the quantum efficiency of energy harvesting systems. A recent study by Paternostro et al. \cite{sgroi2022efficient} has demonstrated how optimizing the local energies of a Fenna-Matthews-Olson complex can achieve highly efficient excitation transfer, even under varying environmental conditions.
A closely related interesting line of research is about keeping coherence properties of the quantum system throughout time (see also \cite{ullah2023preparation}).

Furthermore, the implementation of interacting ancillary systems holds great promise in increasing the efficiency and reliability of quantum algorithms. For instance, as already mentioned, quantum tunneling plays a crucial role in the successful realization of quantum gates. Additionally, it has the potential to contribute to the development of more effective error correction methods, thereby ensuring a more robust preservation of quantum information.
In the context of classical computing and semiconductor devices, the decrease of tunneling probabilities could significantly improve the energy efficiency of energy conversion processes in tunnel diodes and more in general in microelectronics \cite{nano}. Indeed (see Supplementary Material) our approach can be also used to efficiently decrease the tunnelling probability.

In summary, by unlocking the potential for controlling  tunneling probabilities, this work opens the doors for advancements in many fields such as classical and quantum computing, materials science, quantum devices and energy harvesting devices. 

\section*{Methods}

\subsection*{Two-level systems tunneling probability}

The Hamiltonian of a two-level system can always be written in the form 
\cite{Sakurai}
$$
H_S=-\Delta\,\sigma_{z}\, -\, \gamma\, \sigma_{x}=-\hbar\,\omega\,\frac{1+\vec{n}\cdot\vec{\sigma}}{2}\,+\,\hbar\,\omega\,\frac{1-\vec{n}\cdot\vec{\sigma}}{2}\ ,
$$
with $\hbar\,\omega=\sqrt{\Delta^2+\gamma^2}$, $\sigma_{x,y,z}$ the Pauli matrices and 
$\displaystyle\frac{1\pm\vec{n}\cdot\vec{\sigma}}{2}$ its orthogonal eigen-projections, where $\vec{\sigma}=(\sigma_x,\sigma_y,\sigma_z)$ and $\vec{n}=(\gamma,0,\Delta)/(\hbar\,\omega)$. 
The states corresponding to one particle in the left, respectively right well satisfy  
$\sigma_z\ket{L}=-\ket{L}$ and $\sigma_z\ket{R}=\ket{R}$. Starting at $t=0$ in the left well, at time $t$, $\ket{L}$ will evolve into 
$$
\ket{L}_t=\Bigg(\cos(\omega t)\,+\,i\,\vec{n}\cdot\vec{\sigma}\,\sin(\omega t)\Bigg)\ket{L}\ .
$$
Then, the left-to-right tunneling probability $P_{L\to R}(t)$ in~\eqref{1partunn} follows.

\subsection*{Jordan-Schwinger representation}
Two-mode bosons are described by creation and annihilation operators $a, a^\dag$, $b, b^\dag$ satisfying the commutation relations $[a,a^\dag]=[b,b^\dag]=1$, while all other commutators vanish. The operational meaning of $a^\#$ and $b^\#$ is as follows: if $\ket{vac}$ denotes the vacuum state such that $a\ket{vac}=b\ket{vac}=0$, then $a^\dag\ket{vac}$ creates a particle in the left well and $b^\dag\ket{vac}$ a particle in the right one (more details in the Supplementary Material, where the notation $\ket{k} \propto (a^{\dagger})^{k}(b^{\dagger})^{N_S-k} \ket{vac}$ is introduced to indicate the state with $k$ particles in the left well and $N_S-k$ in the right one).  
Using the Jordan-Schwinger representation of the $su(2)$ algebra, we introduce the operators 
\begin{equation*}
J_x \equiv \frac{a^\dag b+ab^\dag}{2}\ ,\ J_y \equiv \frac{a b^\dag-a^\dag b}{2i}\ ,\ J_z \equiv \frac{b^\dag b-a^\dag a}{2}
\end{equation*}
satisfy the algebraic relations of the rotation-group $[J_x,J_y]=iJ_z$, together with $[N_S,J_x]=[N_S,J_y]=[N_S,J_z]=0$. The total number operator of the system is 
$N_S = a^\dag a\,+\,b^\dag b$ and it is left invariant by the rotations generated by $J_{x,y,z}$. 

\subsection*{Beating the $N_S=1$ bound}

As stated in the main text, in the following we adopt $\Delta_S$ as energy unit and $\hbar/\Delta_S$ as time unit.

In the case of a single trapped boson with $N$ trapped ancillary bosons, the left-to-right tunneling probability $P_{L\to R}(t)$ in~\eqref{2qubitprob2} consists of $N+1$  positive terms each bounded by $\displaystyle\frac{\gamma^2}{\omega_k^2}$. In general, it is possible to beat the bound $\displaystyle\frac{\gamma^2}{\omega^2}$ that holds in the case of a single uncoupled trapped boson ($N_A=0$) and can indeed be 
retrieved by switching off the interaction between the two double-well potentials ($\alpha=0$).
As for reaching maximum probability, $P_{L\to R}(t)=1$, notice that the ratio $\displaystyle\frac{\gamma^2}{\omega_k^2}\leq 1$ unless $2=\alpha(N-2k)$. As this can occur for only one $\displaystyle 
k=k^*=\frac{N}{2}-\frac{1}{\alpha}$ at most, unit probability can be reached only for $\ket{\Psi_A}=\ket{k^*}$ at $\displaystyle t_\ell^*=\frac{\pi}{\omega_{k^*}}+\ell\,\pi$, $\ell\in\mathbb{Z}$.

\subsection*{Noisy regime: stationary states}

The right-hand-side of eq. \eqref{GKSL} is the typical generator of the dissipative dynamics of an open quantum system in weak interaction with its environment: it contributes by adding to the commutator with the Hamiltonian two terms, the first one amounting to a particular kind of quantum noise, embodied in the operators $J_z$,  and the second one to a damping term which restores probability preservation.
The choice of the dissipative modification of the reversible dynamics is based upon the fact that it tends to eliminate the off-diagonal terms $\langle k\vert \rho(t)\vert\ell\rangle$ of the time-evolving density matrices which are those sustaining the left-to-right tunneling probability. 
Indeed, the operator $J_z$ has the vectors $\vert k\rangle$ as eigenvectors so that the corresponding eigen-projectors are left invariant 
by the dissipative term:
\begin{equation*}
\label{Noise2}
J_z\,\ket{k}\bra{k}\,J_z-\frac{1}{2}\left\{J_z^2\,,\,\ket{k}\bra{k}\right\}=0\ .
\end{equation*}
Then, the dissipative open dynamics~\eqref{GKSL} is expected to favor states that are mixtures of left/right localized states $\ket{k}\bra{k}$ with weights that depend on the initial state. Notice however that the projectors $\ket{k}\bra{k}$ are not left invariant by the Hamiltonian when it contains a non-vanishing tunneling term ($\gamma\neq 0$). Indeed, the only state of the two-well open system which is left invariant by the master equation~\eqref{GKSL} is the completely mixed state
$\rho_{mix}:=\frac{1}{N_S+1}\sum_{k=0}^{N_S}\ket{k}\bra{k}$ to which all eigen-projectors of $J_z$ equally contribute.
Such a state is the only one which commutes with both a generic Hamiltonian $H_S$ and $J_z$. Therefore, it represents a stationary state of the master equation~\eqref{GKSL} and all initial states $\rho_S$ of the $N_S$ trapped bosons, that evolve into $\rho^{(S)}(t)$ by means of~\eqref{GKSL}, actually tend to it asymptotically in time~\cite{Frigerio78}:
\begin{equation}
\label{limitst}
\lim_{t\to+\infty}\rho^{(S)}(t)=\rho_{mix}\qquad\forall \rho^{(S)}\ .
\end{equation}

\subsection*{Physical resources and experimental feasibility}

\noindent A possible experimental platform to physically implement 
the previously studied model could be provided by dipolar ultracold atoms in optical potentials \cite{BDZ,Lahaye2009,Kawaguchi2012}. The setup would envisages as system an ultracold gas in a double-well potential, realized for instance by combining an harmonic potential and an optical lattice potential \cite{Albiez2005} (see Figure \ref{fig:resources} in the Supplementary Material). One could then  consider as ancilla a spatialy separated ultracold gas, also in a double-well potential. The geometric configuration needed for such a setup could be provided by a ladder configuration, such as that introduced in \cite{Hofferberth2007}. Tuning the parameters of the external potentials for the system and the ancilla, and the distance between the two traps, one could control the parameters described in the text. Estimates for the system and ancilla coefficients $\gamma,\Delta,\eta$ and the coupling $\alpha$ are reported in the Supplementary Material. It turns out that the learned parameters $\gamma_A,\Delta_A,\eta_A$ and $\alpha$ are experimentally feasible.\\

\subsection*{Optimization}

As explained in the main text, our objective is to learn an optimal ancilla Hamiltonian, ancilla initial state, system-ancilla interaction strength and time in order to maximize the probability of 
tunneling the trapped system particles from the left to the right well of the trap.

Concretely, given the density-density interaction $H_{int}=\alpha\, J_{z}\otimes J_{z}$,
the states $\rho^{(SA)}$ of the system plus ancilla evolve in time according to the Hamiltonian
\begin{equation*}
 H_{SA} = H_{S}\otimes  I +  I\otimes H_{A} + H_{int} \ ,
\end{equation*}  
where $I$ denote the identity operator.
For an initial state $\rho^{(SA)}=\rho^{(S)}_{L}\otimes\rho^{(A)}$ with all bosons in the left well of the target system, the probability at time $t$ that $k$ bosons in the system are found in the right well, and thus $N_S-k$ in the left one, is given by 
\begin{equation*}
P_{N_S\to N_{S-k}}(t) = \langle N_S-k\vert\rho^{(S)}_{L}(t)\vert N_S-k\rangle,
\end{equation*}
where the reduced state of the system $S$ at time $t$ is given by~\eqref{red-dyn}.

Our purpose is to maximize the tunneling probability of the $N_S$ bosons in the system $S$, first in the absence of noise and then in the presence of noise.
Towards this goal, we start with an initial state of the system with all particles in the left well and a random initial state of the ancilla.
To maximize the tunneling probability, we parameterized the ancillary Hamiltonian and  interaction strength using the learnable parameters $\eta_{A}, \gamma_{A}, \Delta_{A}$, $\alpha\in \mathbb{R}$ and $t\in\mathbb{R}_{+}$. In specific, we used these parameters to determine:
\begin{equation*}
    \label{ancilla params}
    H_A=\eta_A J_z^2 -\gamma_A J_x - \Delta_A \sigma_z,\ \quad H_{int}=\alpha\,J_z\otimes J_z .
\end{equation*}
We aim to maximize, using automatic gradient techniques, the tunneling probability function
\begin{equation*}
P(\rho_A,\eta_{A},\gamma_{A},\Delta_{A},\alpha,t) = {\rm Tr}\Big(\rho^{(S)}_R \rho^{(S)}_{L}(t)\Big)   
\end{equation*}
where $\rho^{(S)}_R$ is the state of the system with all particles at right, where, in the noiseless case, 
\begin{equation*}
\rho^{(S)}_{L}(t) = {\rm Tr}_{A}\big(e^{-i\,H_{SA}\,t}\,\rho^{(S)}_L\otimes \rho^{(A)} e^{i\,H_{SA}\,t}\big)
\end{equation*}
with  $\rho^{(A)}\in \mathbb{C}^{N_A\times N_A}$ the density matrix of the ancillary system, to be jointly learned with the other parameters $\eta_{A},\gamma_{A},\Delta_{A},\alpha, t$.
In the noisy case dynamics, described by equation \eqref{eq:Noisy}, we utilized  a fourth-order Runge-Kutta method (see \cite{bookRK}, pag. 215). We used this algorithm since  it was  ensuring a stable evolution, compared to a straightforward Euler method. The specific algorithm we employed is as follows: 
\begin{algorithm}[H]
\SetAlgoLined
\KwData{Initial density matrix $\rho(0)$, final time $T$,  steps per unit time $K=int(1/dt)$, Hamiltonian $H$, noise strength $\lambda_A$, $\lambda_S$, time step $dt$, density matrix $\rho$}
\KwResult{$\rho(T)$}
\For{$j \leftarrow 1$ \KwTo $T \times K$}{
    \tcp{Compute the derivative of the density matrix at the current time step}
    $k_1 \leftarrow -i \left( H  \rho - \rho  H \right) + \lambda_A \left( I\otimes J_z  \rho I\otimes J_z - 0.5 \left( I\otimes J_z^2  \rho + \rho  I\otimes J_z^2 \right) \right) + \lambda_S \left( J_z \otimes I  \rho  J_z \otimes I - 0.5 \left( J_z^2 \otimes I \rho + \rho J_z^2 \otimes I \right) \right)$\;
   
    $k_2 \leftarrow -i \left( H  \left( \rho + dt  \frac{k_1}{2} \right) - \left( \rho + dt  \frac{k_1}{2} \right)  \right) + \lambda_A \left( I\otimes J_z  \left( \rho + dt  \frac{k_1}{2} \right) I\otimes J_z - 0.5 \left( I\otimes J_z^2  \left( \rho + dt  \frac{k_1}{2} \right) + \left( \rho + dt  \frac{k_1}{2} \right)  I\otimes J_z^2 \right) \right) + \lambda_S \left( J_z\otimes I  \left( \rho + dt  \frac{k_1}{2} \right)  J_z\otimes I - 0.5 \left(  J_z^2 \otimes I  \left( \rho + dt  \frac{k_1}{2} \right) + \left( \rho + dt  \frac{k_1}{2} \right)  J_z^2 \otimes I \right) \right)$\;

    $k_3 \leftarrow -i \left( H \left( \rho + dt  \frac{k_2}{2} \right) - \left( \rho + dt  \frac{k_2}{2} \right)  H \right) + \lambda_A \left( I\otimes J_z  \left( \rho + dt \frac{k_2}{2} \right)  I\otimes J_z - 0.5 \left( I\otimes J_z^2 \left( \rho + dt \frac{k_2}{2} \right) + \left( \rho + dt \frac{k_2}{2} \right)  I\otimes J_z^2 \right) \right) + \lambda_S \left( J_z \otimes I  \left( \rho + dt \frac{k_2}{2} \right)  J_z \otimes I - 0.5 \left( J_z^2\otimes I  \left( \rho + dt \frac{k_2}{2} \right) + \left( \rho + dt  \frac{k_2}{2} \right) J_z^2 \otimes I \right) \right)$\;

    $k_4 \leftarrow -i \left( H  \left( \rho + dt  k_3 \right) - \left( \rho + dt  k_3 \right)  H \right) + \lambda_A \left( I\otimes J_z \left( \rho + dt k_3 \right) I\otimes J_z - 0.5 \left( I\otimes J_z^2  \left( \rho + dt  k_3 \right) + \left( \rho + dt k_3 \right)  I\otimes J_z^2 \right) \right) + \lambda_S \left(J_z\otimes I \left( \rho + dt  k_3 \right)  J_z \otimes I  - 0.5 \left( J_z^2 \otimes I \left( \rho + dt  k_3 \right) + \left( \rho + dt  k_3 \right)  J_z^2 \otimes I  \right) \right)$\;
    \tcp{Update the density matrix at the current time step}

    $\rho \leftarrow \rho + \frac{dt}{6} \left( k_1 + 2 k_2 + 2  k_3 + k_4 \right)Heaviside(\widehat{t}-jdt)$\;   
}
\caption{Noisy case Evolution Algorithm}\label{algo}
\end{algorithm}
Importantly, the $\widehat{t}$ variable is also optimized by automatic differentiation within the time window $[0,T]$. This is achieved by multiplying the $\rho$ update at step $j$ by the function $Heaviside(\widehat{t}-jdt)$ which eliminates the updates after time $\widehat{t}$.
In concrete the maximization of the tunneling probability $P$ is performed by defining an equivalent minimization problem introducing the loss function $\mathcal{L}:\mathbb{C}^{N_A\times N_A}\times \mathbb{R}^{5}\to \mathbb{R}$ with
$\mathcal{L} = 1-P$. 
In the simplified case of eq. \eqref{2qubitprob2sym} the Loss was added with an additional term penalizing the time, $\lambda \|t\|^2$ with $\lambda=0.1$. 
We employed a very effective and widely used optimizer in machine learning, ADAM \cite{KingBa15}, with a learning rate $lr=0.01$ and automatic differentiation in PyTorch \cite{automaticdiff} a machine learning library of the Python programming language \cite{PaszkePytorch}. 
The learnable parameters $\eta_{A},\gamma_{A},\Delta_{A},\alpha$ are initialized to $1$  without sign constraints during the optimization. The time $t$ is also initialized to $1$ and remains positive during learning. Changing the initialization randomly in the fixed range $[0,1]$ did not change the results. The ancilla $\rho^{(A)}$ is initialized to a random density matrix. The number of iterations was chosen to guarantee the convergence of all learned parameters.
After learning, the optimized parameters are utilized to construct $H_A$ and $H_{int}$ and consequently $H_{SA}$. The Hamiltonians $H_{SA},H_S$ are then employed to produce the plots of the tunneling probability evolution over time using the same evolution functions as during the learning process (but with fixed parameters and no Heaviside function).
The evolution for the noisy system alone (no ancilla) is obtained with the same Runge-Kutta algorithm (Algo \ref{algo}) with $\lambda_A=0$ and $J_z\otimes I \rightarrow J_z$.
This approach allows us to observe and analyze the behavior of the system's tunneling probability over time, based on the optimized parameters, providing valuable insights into the effectiveness of the learning process and the achieved probability improvements.
We also tried to fully learn the interaction Hamiltonian:
\begin{equation*}
H_{int} = \sum_{ij}\alpha_{ij} J_{i}\otimes J_{j}.    
\end{equation*}
No significant improvement was obtained with this strategy.

\section*{Acknowledgments}
F.B., A.dO. and A.T. acknowledge financial support from the PNRR PE National Quantum Science and Technology Institute (PE0000023).

\bibliography{refs_tunneling} 

\newpage

\section*{Supplementary Material}

\subsection*{N-bosons in a double-well potential}

In the following, we shall be concerned with a typical ultracold atom experimental setup consisting in a double-well potential confining $N$ particles of bosonic type described by creation and annihilation operators $a, a^\dag$, $b, b^\dag$ satisfying the commutation relations $[a,a^\dag]=[b,b^\dag]=1$, while all other commutators vanish.
If $\ket{vac}$ denotes the vacuum state such that $a\ket{vac}=b\ket{vac}=0$, then $a^\dag\ket{vac}$ creates a particle in the left well and $b^\dag\ket{vac}$ a particle in the right one. It follows that states with $0\leq k\leq N$ particles in the left well together with $N-k$ in the other one are represented by:
\begin{equation}
\label{ONB1SM}
\ket{k}=\frac{(a^{\dagger})^{k}(b^{\dagger})^{N-k}}{\sqrt{k!\,(N-k)!}} \ket{vac},\;\;\;k=1,\cdots,N\ .  
\end{equation}
Notice indeed that these vectors fulfill 
\begin{equation}
\label{ONB2}
a^\dag b\ket{k}=\sqrt{(k+1)(N-k)}\,\ket{k+1}\ ,\quad ab^\dag\ket{k}=\sqrt{k(N-k+1)}\,\ket{k-1}
\end{equation}
and are thus eigenstates of the number operators $a^\dag a$ and $b^\dag b$:
\begin{equation}
\label{ONB3}
 a^\dag\,a\,\ket{k}\,=\,k\,\ket{k}\ ,\quad b^\dag\,b\,\ket{k}\,=\,(N-k)\,\ket{k}\ .
\end{equation}
As such, they constitute an orthonormal basis for the Hilbert space $\mathbb{C}^{N+1}$ associated with the system $S$ just described. Moreover,
in the Jordan-Schwinger representation of the $su(2)$ algebra, the operators 
\begin{equation}
J_x \equiv \frac{a^\dag b+ab^\dag}{2}\ ,\ J_y \equiv \frac{a b^\dag-a^\dag b}{2i}\ ,\ J_z \equiv \frac{b^\dag b-a^\dag a}{2}
\end{equation}
satisfying the algebraic relations proper to the generators of the rotation group:
\begin{equation}
 [J_x,J_y]=\,i\,J_z\ ,
 \end{equation}
 plus cyclic permutations, together with
 \begin{equation}
 [N,J_x]=[N,J_y]=[N,J_z]=0\ 
\end{equation}
and $N$ denoting the total number operator 
\begin{equation}
\label{numberop}
 N \equiv a^\dag a\,+\,b^\dag b\ .  
\end{equation}
Furthermore, their matrix elements with respect to the ONB~(\ref{ONB1SM}) are
\begin{eqnarray}
 \label{JS2a}  
 \langle j\vert J_x\vert k\rangle&=&\frac{\sqrt{(k+1)(N-k)}\,\delta_{j,k+1}+\sqrt{k\,(N-k+1)}\,\delta_{j,k-1}}{2}\\
 \label{JS2b}
  \langle j\vert J_y\vert k\rangle&=&\frac{\sqrt{k\,(N-k+1)}\,\delta_{j,k-1}-\sqrt{(k+1)(N-k)}\,\delta_{j,k+1}}{2i}\\
  \label{JS2c}
   \langle j\vert J_z\vert k\rangle&=&\frac{N-2k}{2}\delta_{j,k}\ .
\end{eqnarray}

\subsection*{Two simple limit cases: details}
As a benchmark simple example in the noiseless setting, consider a
double-well containing one boson, $N_S=1$, coupled to an ancillary double-well with $N_A \equiv N$ non-interacting and non-tunneling bosons: 
\begin{equation}
    H_S=-\Delta\,\sigma_z-\gamma\,\sigma_x\ ,\quad H_A=-\Delta\,J_z\ ,\quad H_{int}=\alpha\,\sigma_z\otimes J_z\ .
\end{equation}
By means of the orthogonal eigen-projections $P_k=\ket{k}\bra{k}$ of $J_z$ such that 
$$
\sum_{k=0}^NP_k=
I\ ,\quad J_z=\sum_{k=0}^N\frac{N-2k}{2}\,P_k\ ,
$$
one writes, neglecting terms proportional to the identity,
\begin{equation}
    \label{2qubitHam}
    H_{SA}=\sum_{k=0}^N\,H_k\otimes P_k\ ,\qquad H_k:=-\left(\Delta-\alpha\frac{N-2k}{2}\right)\,\sigma_z-\gamma\sigma_x\ .
\end{equation}
Then, the unitary time-evolution  generated by $H_{SA}$ reads
\begin{equation}
    \label{2qubitdyn}
    {\rm e}^{-i\,t\, H_{SA}}=\sum_{k=0}^N{\rm e}^{-i\,t\,H_k}\otimes P_k\ .
\end{equation}
Consider the joint initial state $\rho^{(SA)}$ of system and ancilla to be a tensor product state $\rho^{(SA)}=\rho^{(S)}_L\otimes\rho^{(A)}$ corresponding to the system with its single boson localized in the left well, $\rho^{(S)}_L=P_N$ and to the ancilla being in a generic pure state $\rho^{(A)}=\ket{\psi_A}\bra{\psi_A}$ of its $N$ bosons. According to~\eqref{red-dyn}, the system's initial state projector evolves into 
\begin{equation}
    \label{2qubitdynvec}
  \rho^{(S)}_L(t)=\sum_{k=0}^N\,\Big|\langle k\vert\psi_A\rangle\Big|^2\,{\rm e}^{-i\,t\,H_k}\,\ket{L}\bra{L}\,{\rm e}^{i\,t\,H_k}\ .
\end{equation}
Therefore, the left-to-right transition probability at time $t\geq 0$ amounts to
\begin{equation}
    \label{2qubitprob}
P_{L\to R}(t) = \langle R\vert\rho^{(S)}_L(t)\vert R\rangle=\sum_{k=0}^N\,\Big|\langle k\vert \psi_A\rangle\Big|^2\, 
\Big|\langle R\vert {\rm e}^{-i\,t\,H_k}\vert L\rangle\Big|^2\ ,
\end{equation}
where $\ket{R}$ is the state with all system bosons to the right.
Using~\eqref{1partunn} with $H_k$ as in~\eqref{2qubitHam}, one finally finds
\begin{equation}
    \label{2qubitprob2bis}
P_{L\to R}(t) = \sum_{k=0}^N\Big|\langle k\vert\psi_A\rangle\Big|^2\,\frac{\gamma^2}{\omega_k^2}\,\sin^2(\omega_k t)\ ,\quad\omega_k:=\sqrt{\left(\Delta-\alpha\frac{N-2k}{2}\right)^2\,+\,\gamma^2}\ ,
\end{equation}
from which the expression~\eqref{2qubitprob2} of the main text follows setting $\Delta=1$.

For the second simple example with the Hamiltonians in~\eqref{2qubitssymm}, the argument is the same as above, only the Hamiltonians $H_k$ read, in unit of $\gamma_S$,
\begin{equation}
    \label{2qubitHamsymm}
H_k:=-\gamma_k\,\sigma_x\ ,\qquad \gamma_k:=1-\alpha\frac{N-2k}{2}\ ,
\end{equation}
thus yielding the tunneling probability in~\eqref{2qubitprob2sym}.

\subsection*{Experimental parameters}
Let us consider the following form for the potential $V(x,y,z)$ for the ultracold system as the sum of a contribution of a (magnetic) harmonic potential $V_{h}$ and 
of an optical lattice potential $V_l$ along the $x$-direction:
$
V(x,y,z)=V_h(x,y,z)+V_l(x)\, ,$
where the harmonic potential reads $V_h(x,y,z)=\frac{m}{2} \left[ \omega_x^2 x^2 + \omega_r^2 (y^2+z^2) \right)$ and the periodic potential $V_l(x)=V_0 \cos^2(kx)$. Typically (\cite{Morsch2006}) one has $k = 2 \pi/\lambda$, where $\lambda = \lambda_{laser} \sin (\theta/2)$, with $\lambda_{laser}$ 
being the wavelength of the lasers giving raise to the optical potential and $\theta$ the angle between
the counterpropagating laser beams (we consider simply $\theta=\pi$ in the following). Energies are typically expressed in units of the recoil energy $E_R=\hbar^2 k^2 / 2 m$, $m$ begin the mass of the atom. To have a one-dimensional geometry one needs to have $\omega_r \gg \omega_x$. By varying the parameters $\omega_x$ and $V_0$ one can have double- or multi- well configurations \cite{Albiez2005,Macri2013}. Moreover, 
with $V_l(x)=V_0 \cos^2(kx + \phi)$, by varying the phase $\phi$ (or equivalently by shifting the center of the $x$-part of the harmonic potential $V_h$), one can induce an energy imbalance $\Delta E$ between the adjacent wells.  

The coefficient $\gamma_S$ in \eqref{HamS} is proportional to the tunneling coefficient, typically denoted by $K$, of the two-mode model \cite{Smerzi97}. One can perform a very simple estimate of $K/E_R$ by considering a variational ansatz of the (Wannier) spatial wavefunctions localized in the well \cite{Trombettoni2005}. Putting $V_0=s E_R$ and using $\lambda \sim 800 nm$, one finds for (bosonic) $Yb$ atoms $\gamma_S \sim k_B 0.5 nK$ with $s=10$. The coefficient $\Delta_S$ in \eqref{HamS} depends on the energy asymmetry $\Delta E$: with $\omega_x \sim 2 \pi 100 Hz$ and again $s=10$, one gets $\Delta_S \sim 0.2 nK$ for $\phi=0.5$. The coefficient $\eta_S$ is proportional to the interaction coefficient, usually denoted by $U$, of the two-mode model. It depends on the transverse size, i.e. in the $-z$ directions of the system and of course on the number of particle \cite{Smerzi03,Ananikian06}: one finds with $\sigma_\perp$ of few microns values from fraction to some $nK$ with typical values \cite{Cataliotti01,Morsch2006}. 

Similar estimates can be done for the coefficients of the ancilla system. Finally the coefficient $\alpha$ of the coupling is the one in front the of the $z-z$ interaction and clearly depends on the distance between the system and the ancilla. A simple estimate can be obtained by writing the localized wavefunction of the system, say $\Phi_S(\vec{r})$, and of the ancilla, say $\Phi_A(\vec{r})$, of course respectively centered in the wells of the system of the ancilla. Then one has $\alpha \propto \int d\vec{r} d\vec{r}'\Phi_S(\vec{r}) {\cal V}(\vec{r}-\vec{r}') \Phi_A(\vec{r}')$, 
where ${\cal V}$ is the non-local potential. A discussion of it for dipolar and Rydberg atoms can be found in \cite{Lahaye2009,Defenu23}. When the angle between the dipoles is $\pi/2$, one can write the $2$-body potential in the form 
${\cal V}(\vec{r}-\vec{r}')= {\cal C}/|\vec{r}-\vec{r}'|^3$. One can tune the distance between the system and the ancilla, and crucially as well the coefficient ${\cal C}$, therefore making possible the tuning of the coefficient $\alpha$.

To conclude this Section, we observe that in the main text, when the energies are expressed in units of $\gamma_S$ and time in units of $\hbar/\gamma_S$ , with $\gamma_S \sim 0.5 nK$, one has that the unit of time is $\sim 10 ms$, so hundreds of oscillations corresponds to a total time of the experiment of order up to few seconds. Similar estimates hold when the energies are measured in units of $\Delta_S$.

\begin{figure}[h]
\centering
\includegraphics[width=0.4\linewidth]{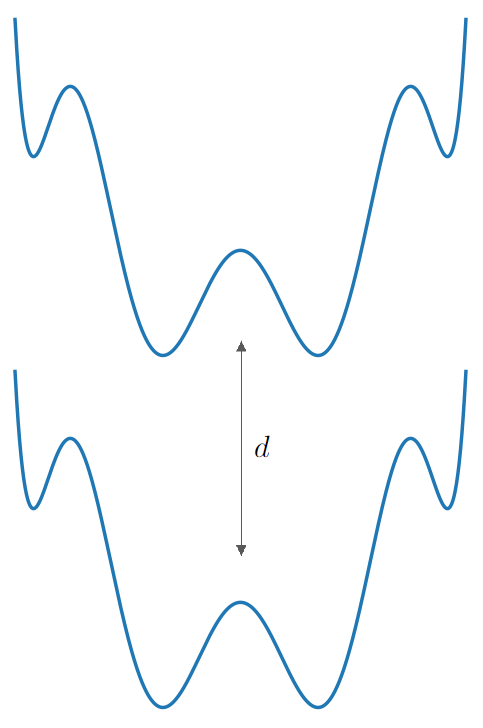}
\caption{Schematic of system-ancilla tuning in a ultracold system of e.g. dipolar and Rydberg atoms.}
\label{fig:resources}
\end{figure}

\subsection*{Supplementary figures}

\subsubsection*{- Ancilla different initialization}
\begin{figure}[h]
\centering
\begin{minipage}{.5\textwidth}
  \centering
\includegraphics[width=1.0\linewidth]{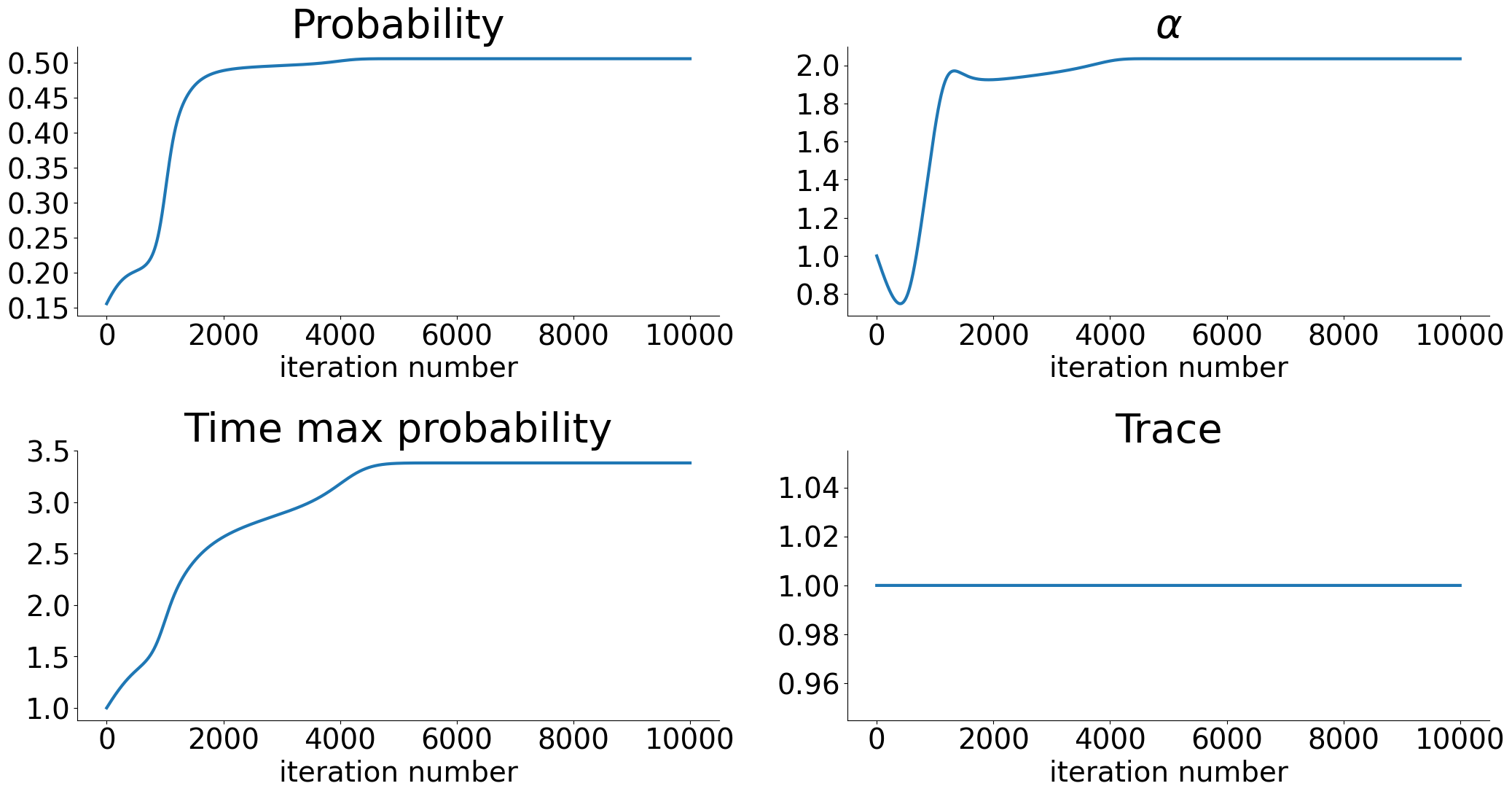}
\end{minipage}%
\begin{minipage}{.5\textwidth}
  \centering
\includegraphics[width=.7\linewidth]{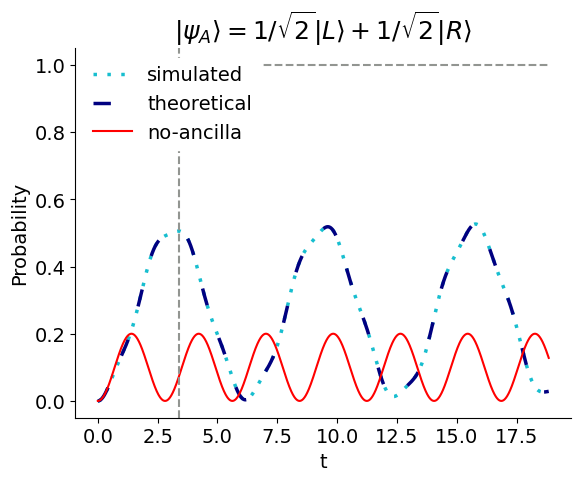}
\end{minipage}
\caption{\textit{Left} Evolution, during optimization, of the parameters for $\eta_S=0$, $\gamma_S=1$, $\Delta_S=1$. \textit{Right} Time evolution of tunneling probability for the 1-boson system coupled with the 1-boson ancilla compared to the system without the ancilla, reporting both the theoretical predictions and the experimental results (for $| \psi_A \rangle = 1/ \sqrt{2}(|L \rangle + |R \rangle)$).}
\label{fig:test2}
\end{figure}

Figure \ref{fig:test2} shows the same plots as in Figure \ref{fig:test1} in the main text but when the ancilla is fixed to a superposed state of right and left one particle during the optimization. The results show full agreement with the theoretical predictions. Interestingly the reached max probability is not one, highlighting the importance of the ancilla initialization.

\subsubsection*{- Further results on the noiseless case}

\begin{figure}
\centering
\begin{minipage}{.45\textwidth}
  \centering
  \includegraphics[width=0.89\linewidth]{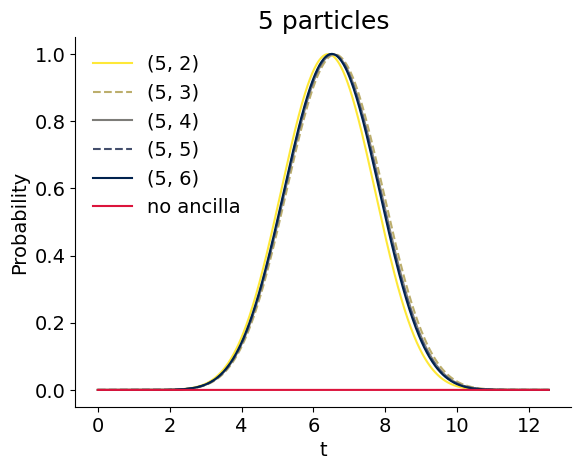}
\end{minipage}%
\begin{minipage}{.45\textwidth}
  \centering
  \includegraphics[width=.9\linewidth]{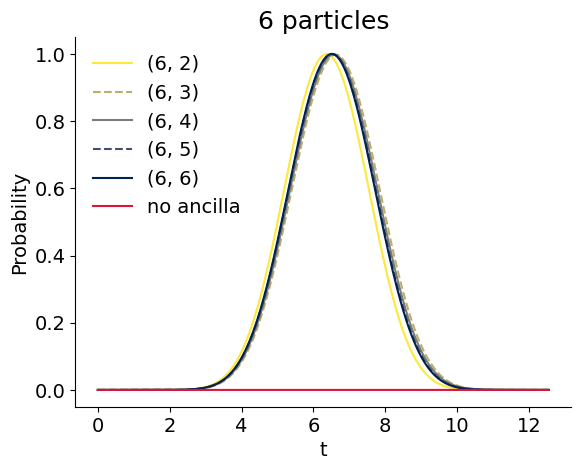}
\end{minipage}
\caption{Time evolution of tunneling probability for $(N_S,N_A)$ particles in the system and ancilla with $N_S=5,6$ and $N_A=2,\cdots,6$ compared with that of the system with no coupling (red line).}
\label{fig:test56}
\end{figure}

Figure \ref{fig:test56} extends the results of Figure \ref{fig:test34} in the main text for the case of $N_S=5,6$ particles in the system. 

\subsubsection*{- Further results on the noisy case}

\begin{figure}[h!]
\centering
\begin{minipage}{.5\textwidth}
  \centering
\includegraphics[width=1.0\linewidth]{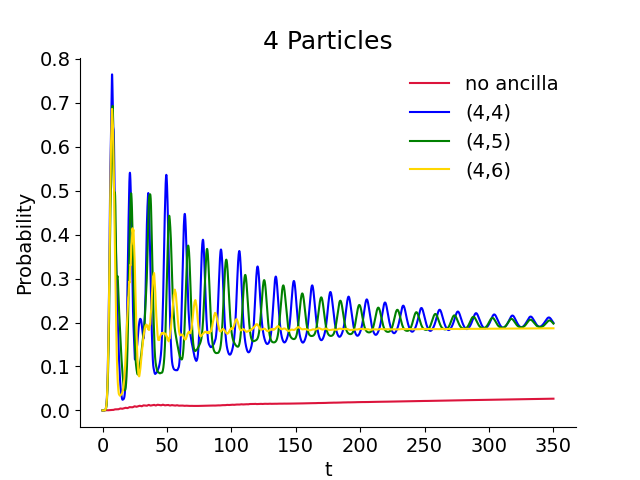}
\end{minipage}%
\begin{minipage}{.5\textwidth}
  \centering
\includegraphics[width=1.0\linewidth]{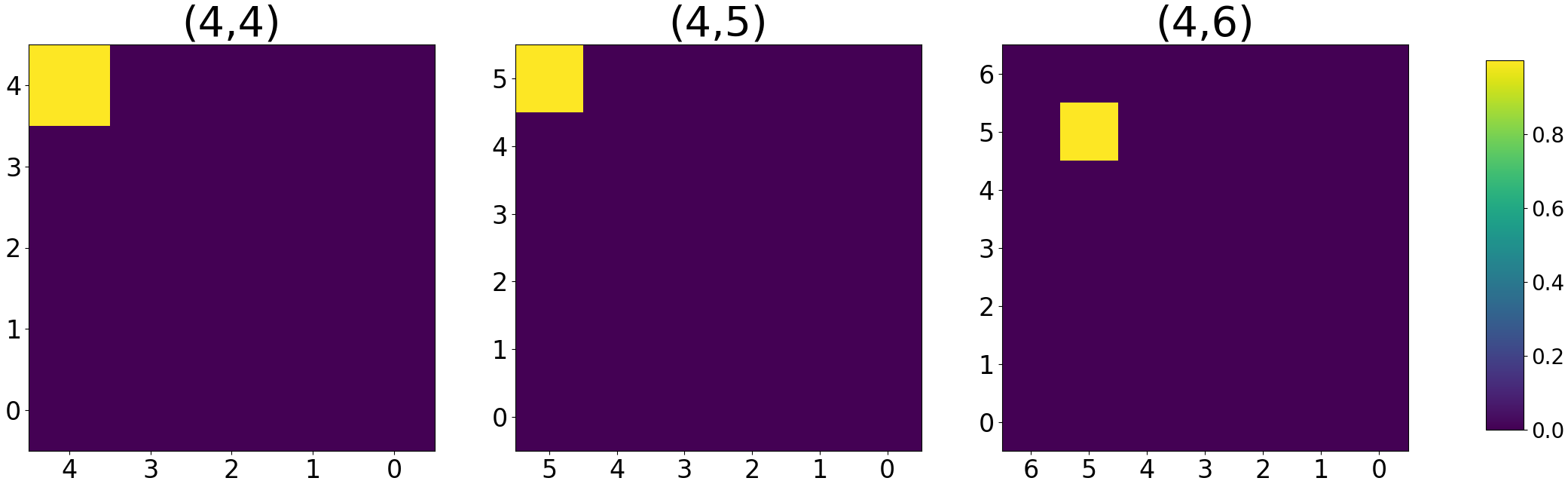}
\end{minipage}
\caption{\textit{Left:} Tunneling probability evolution after optimization for a system of $4$ particles coupled with an ancilla of $4,5,6$ particles. Red line: system with no coupling. Note the difference is the time scales between the right and left plots. Both system and ancilla are initialized, before learning, to the state where all particles are in the left well.\textit{Right:} The learned ancillas.}
\label{fig:ancillas34}
\end{figure}

Figure \ref{fig:ancillas34} (Left) reports the evolution of the tunneling probabilities for $N_S=4$ as in Figure \ref{fig:test34} in the main text but with the constraint of diagonal ancillas. Interestingly, although not explicitly imposed in the learning, the ancilla state converges not to a superposition state of particles but to a state with a precise number of particles in the right and left well. 

The constraint on the learned ancilla has been imposed as follow: at each step of the optimization we extracted the diagonal elements from $\rho_A$, generated a new ancilla density matrix with those diagonal elements and then normalized the result. In specific at each iteration step we perform: 
\begin{align*}
    \rho^{(A)} &\leftarrow \text{diag}(\rho^{(A)}) \\
    \rho^{(A)} &\leftarrow \frac{\rho^{(A)}}{{\rm Tr}(\rho^{(A)})}.
\end{align*}

\subsection*{- Decreasing the tunneling probability}
\begin{figure}
\centering
  \includegraphics[width=0.6\linewidth]{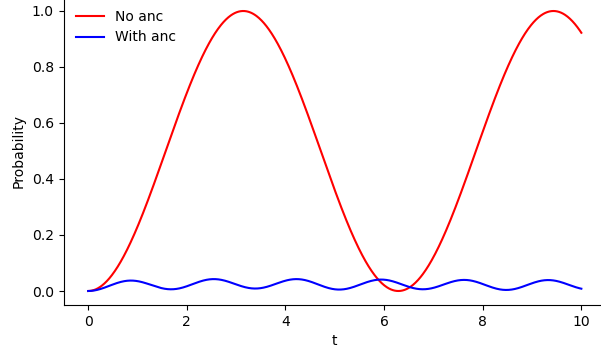}
\caption{Tunneling probability evolution after optimization for a system of $1$ particles coupled with an ancilla of $6$ particles. Red line: system with no coupling.}
\label{fig:pdecrease}
\end{figure}

Interestingly, as shown in Figure \ref{fig:pdecrease}, the same strategy can be applied to decrease the tunnelling probability. Below the case of one particle with $\Delta_{S}=0$ and $\gamma_{S}=1$.  

\end{document}